# Roman CCS White Paper

# Title: Detection and characterization of M-L-T-Y dwarfs belonging to the Milky Way Disks and Stellar Halo with the Roman Space Telescope

**Roman Core Community Survey:** *High Latitude Wide Area Survey, High Latitude Time Domain Survey*

**Scientific Categories:** *stellar populations and the interstellar medium; galaxies*

**Submitting Author:**
Name: Dr. B.W. Holwerda
Affiliation: University of Louisville
Email: benne.holwerda@louisville.edu

**List of contributing authors** (including affiliation and email):
Nor Pirzkal (STSCI, pirzkal@stsci.edu)
Adam Burgasser (UCSD, aburgasser@ucsd.edu)
Chih-Chun Hsu (Center for Interdisciplinary Exploration and Research in Astrophysics (CIERA), Northwestern University, chsu@northwestern.edu)

**Abstract:**

How many low-mass stars, brown dwarfs and free-floating planets are in the Milky Way? And how are they distributed in our Galaxy? Recent studies of Milky Way interlopers in high-redshift observations have revealed a 150-300 pc thick disk of these cool stars with 7% of the M-dwarfs in an oblate stellar halo. One can use the High Latitude Survey with the Roman Space Telescope to search for Galactic ultracool dwarfs (spectral classes M, L, T, and Y) to accurately model the 3D structure and the temperature and chemical evolution of the Milky Way disk in these low-mass (sub)stellar objects.

Accurate typing has been shown to work on HST grism and photometric data using machine learning techniques. Such an approach can also be applied to Roman photometry, producing accurate photometric typing to within two subtypes. The High Latitude Survey provides enough statistical power to model the Milky Way structural components (thin and thick disks and halo) for M-, L- and T/Y-dwarfs. This approach has the benefit to allow us to constrain scale-lengths, scale-heights and densities, as well as the relative position of our Sun with respect to the disk of dwarf stars of our Milky Way. The total number of each brown dwarf type can be used to infer both the low-mass end of the Galaxy-wide Initial Mass Function (IMF) for the first time, the formation history of low-mass stellar and substellar objects, and the fraction of low-mass stars in the halo, a statistic that can test cold dark matter structure formation theories.

**Delving for Dwarfs at High Latitude**

Low-mass stars do not die, they merely fade away. All the brown dwarfs the Milky Way has ever produced (or accreted) must reside somewhere in the Galactic disk and halo. Roman Space Telescope observations are extremely well suited to find and map the brown dwarfs of the Milky Way. An accurate tally of low-mass stars and brown dwarfs of the Galaxy is critical to understand the Initial Mass Function (IMF), the number of free-floating planets, and the stellar mass budget of our Galaxy.

Counting stars to infer the shape and size of our Milky Way Galaxy is a classic experiment in astronomy (e.g., Herschel 1785; Kapteyn, 1922). Inferring the shape from star counts is prone to observational bias and sampling issues, both now well-understood. Observational interest has moved from massive and solar-type stars (Gilmore & Reid, 1983; Gilmore, 1984; Siegel et al., 2002; Bovy et al., 2012) to low-mass stars, both because these are interlopers in high-redshift galaxy searches (Caballero et al., 2008; Wilkins et al., 2014) and constrain the upper mass limit of free-floating planets (Deacon, 2018). Early efforts used HST deep extra-galactic observations (high Galactic latitude) to constrain M-L-T dwarf star Galactic scale-heights (Ryan et al., 2005; Stanway et al., 2008; Pirzkal et al., 2005, 2009) but these were limited by statistics, a limited number of lines-of-sight, and challenges in photometric or grism classification. These narrow-field space-based surveys are now reaching their limits in WISPS grism classifications, with priors using local kinematics (see Hsu et al., 2021; Ryan et al. 2022; Aganze et al., 2022a,b

Brown dwarfs cool as they age, moving towards later spectral type (e.g. from late M to L). Ryan et al. (2017, 2022) used this behavior and the fact that the population of M/L/T/Y dwarfs is constantly kinematically heated to show that the scale-height of different dwarf star types is directly linked to the star-formation history of the Milky Way (Burgasser et al., 2015; Hsu et al., 2021; Aganze et al., 2022a,b). The vertical distribution of M/L/T/Y dwarf stars thus directly connects to their ages and thus the rate at which stars have been formed in the Milky Way disk and the relative numbers of low-mass stars; the low-mass end of the Galaxy-wide Initial Mass Function (IMF). Roman Space Telescope observations are extremely well-suited to map ultracool dwarf stars because (a) Roman will observe in the near-infrared, (b) the community surveys cover a wide area of the sky at high Galactic latitude, and (c) the filter suite is well suited to photometrically characterize their types.

However, to realize the full potential of Galactic science with Roman, one needs accurately type M/L/T/Y dwarfs spread over a large volume in the disk and model their vertical distribution to large distances (> kpc). 2MASS, WISE and GAIA only probe dwarf stars in the immediate Solar surroundings (~100pc) and therefore provide a very limited view of the Milky Way disk (see e.g. Carnero Rosell et al., 2019; Ahmed & Warren, 2019). HST/WFC3 grism observations can type these objects more accurately but these are still limited in depth to the Milky Way disk (~400 pc for T dwarfs, Aganze et al., 2022a,b).

Hubble has already probed ultracool dwarfs beyond the disk of the Milky Way along a multitude of sight-lines in the near-infrared with pure-parallel WFC3 imaging. These data have already been used in several studies (cf Ryan et al., 2011; Holwerda et al., 2014; van Vledder et al., 2016). The challenge with HST data is to identify and type these stars from their photometry, and model their distribution in space from limited fields of view across the sky. Roman will perform better than these surveys by virtue of more near-infrared filters and much greater field of view.

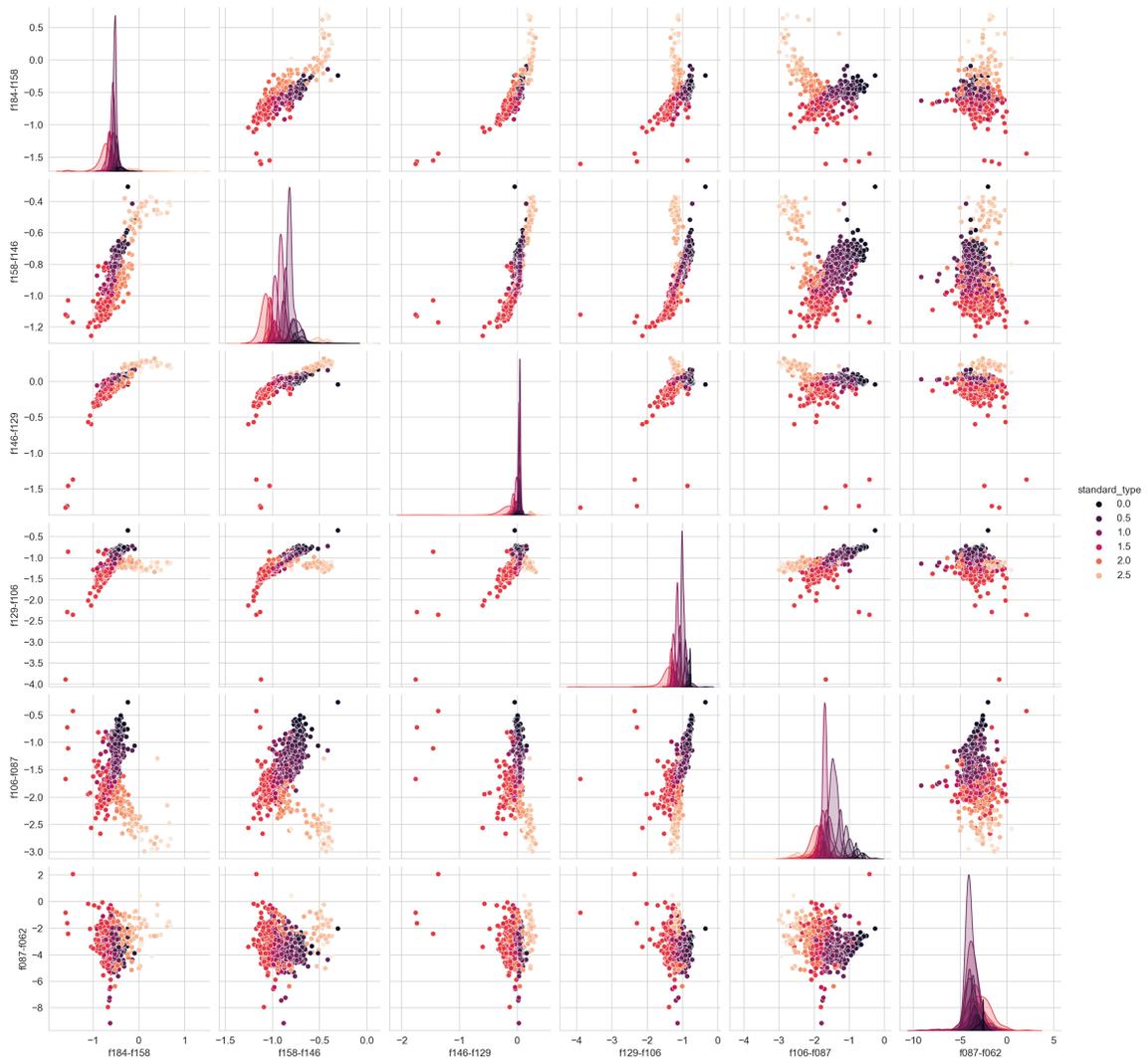

**Figure 1** — The broad filter colors of the Roman Space Telescope predicted using SPLAT spectra (Burgasser & Splat Development Team, 2017). Colors are in AB magnitude and separate out the broad brown dwarf types well (M=0, L=1, T=2, Y=3). New techniques are needed to refine the typing using this photometry.

**Photo-Typing M/L/T/Y objects**

Broad colors can give a reasonable estimate of whether an object is an M-, L-, T- or Y-type ultracool dwarf (see Ryan et al., 2011; Holwerda et al., 2018). However, to more precisely subtype and therefore obtain an accurate estimate of the expected absolute magnitude, one needs more than just a single broad color. Some studies have used grism observations (Pirzkal et al., 2005, 2009; Aganze et al., 2022a,b) or proper motion (Kilic et al., 2005). What is needed is sufficient information around 1 micron to accurately type dwarfs, an area where the Roman High Latitude and the variability surveys will excel. The wide Roman NIR filters colors (Figure 1) combined with machine learning techniques such as k-nearest neighbors (kNN), one can already type to within a few subtypes (e.g. M2±2 vs M6±2 etc. see Fig. 2).

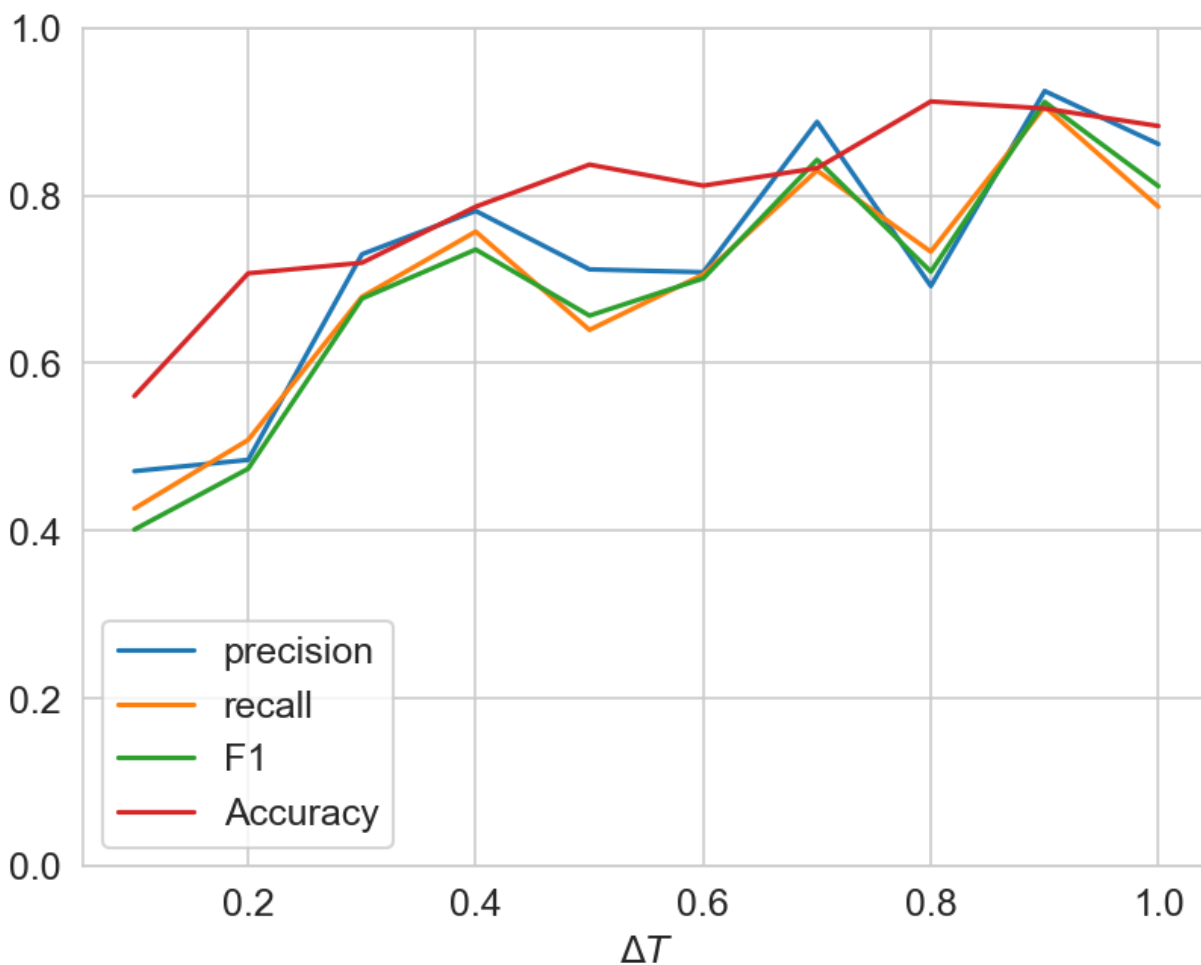

**Figure 2** — the machine learning metrics of the k-nearest neighbor performance as a function of the desired type resolution ($\Delta T$). Type resolution is the width of the bin into which the kNN algorithm places an object. The training set is Roman filters calculated from SPLAT spectra (Burgasser & Splat Development Team, 2017). A reasonable performance (close to 80% precision, recall and F1=precision x recall / precision+recall) can be achieved within 2 subtypes ($\Delta T = 0.4$, a bin 4 subtypes wide).

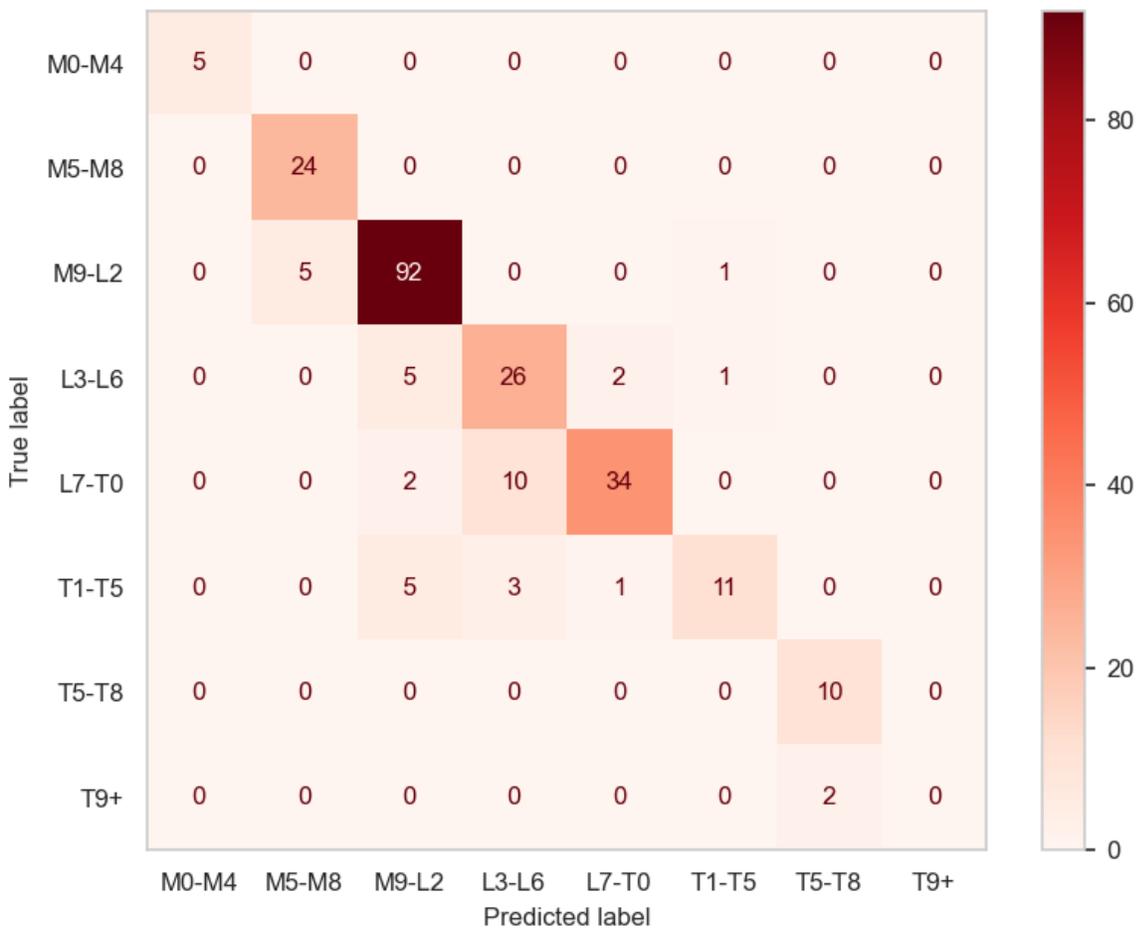

**Figure 3 —** the performance of the kNN algorithm trained on Roman Space Telescope NIR colors calculated from SPLAT spectra (Burgasser & Splat Development Team, 2017). Classification is already quite good to within a few subtypes. Misclassifications tend to be towards earlier types and offer clues to possible refinements. Yet this level of accuracy would allow for a myriad of Milky Way structure questions to be answered with the Roman High Attitude Survey. The training set can be boosted by observations of known ultracool dwarfs in the survey data.

We note that the bluest and reddest filter (F062W and F213W respectively) are not fully sampled in the Spex spectra (SPLAT, Burgasser & Splat Development Team, 2017; Holwerda et al., 2018). A wide suite of atmospheric models with enough variance and statistics may well increase the accuracy of an otherwise simple kNN approach (See Figure 3).

**Modeling the Milky Way**

In order to improve our understanding of Galactic structure for brown dwarfs, improvements in statistics, photometry, and number of sight-lines are critical. At present, the deep pencil-beam searches for faint, high-redshift galaxies with HST/WFC3 and JWST/NIRcam constitute the best existing data-sets to search for these objects outside the immediate Solar neighborhood (much better-suited WISE data exists for stars within <100 pc). However, the accuracy on any measurements of Galactic properties is very much limited by the number of lines-of-sight, not just the raw statistics along one line-of-sight. Additional prior constraints come from the local

IMF ([Kirkpatrick et al., 2021](#)), local 3D kinematics ([Hsu et al., 2021](#)), and the previous WISPS scale-heights ([Aganze et al., 2022b](#)). Roman promises to deliver statistics and lines of sight through the Milky Way on a grand scale. One can image not only identifying the broad structure of the Milky Way but also deviations, such as streams of low-mass stars and ultracool dwarfs in the halo.

**Science Questions**

With this much improved Roman Survey data in hand, the following science questions can be addressed:

1. Does Milky Structure depend on brown dwarf class? Using different probes of the distribution of brown dwarfs (e.g., M- vs L-dwarfs), what are the typical scales of the disk and halo? Are they consistent or is there a structural dependence as there is with stellar metallicity and age (see e.g. [Bovy et al., 2012](#))? Can we distinguish the effects of cooling and star-formation history on the brown dwarf population? ([Ryan et al., 2017, 2022](#))?

2. How prominent is the thick disk for brown dwarfs? In [van Vledder et al. (2016)](#), this component was not included. In [Hsu et al., 2021](#) it was already evident with anecdotal evidence in photometry ([Schneider et al. 2020](#)). The prominence of this second disk remains somewhat contentious ([Bovy et al., 2012](#)). Searches for thick disk and halo objects will help us identify metal-poor ultracool dwarfs for spectroscopic follow-up studies.

3. What is the scale-length of the Milky Way for brown dwarfs? Previously, this variable was fixed due to the limited range of Galactic radii sampled in Galactic latitude fields. Both large-scale imaging and better typing, will allow us independently constrain the scale-length for low-mass objects.

4. What is the total number of M, L T & Y dwarfs in the Milky Way? Integrating the best MCMC model of each (sub)type will result in total numbers of brown dwarf per type and Galactic structural component. The ratio between these is effectively the low-mass end of the Galaxy-wide IMF.

5. Position of the Sun – [Pirzkal et al. (2009)](#) showed a North/South discrepancy between M-dwarf counts. This may point to a misaligned brown dwarf disk with other stellar types.

**Observing Strategy**

The current observing strategy should be sufficient (dithers to cover chip gaps and dead pixels) with enough spatial resolution to identify unresolved sources reliably to about a magnitude above the survey depth. Multiepoch data would aid in identification of Galactic origin through proper motion.

**Filter Choice**

The filter choice of Y106/J129/H158/F184 for the High Latitude Survey is sufficient for sub typing to within two subtypes, and that is likely sufficient to map Milky Way substructure as a function of type. More detailed models down to subtype precision may need grism observations of these objects.

**Summary:**

The Roman Space Telescope imaging capability offers already a powerful tool to quantify the numbers of brown dwarfs residing in multiple Milky Way structural components (thin disk, thick disk, stellar halo, bulge, and possibly streams). The filter suite can identify low-mass and sub-stellar objects to within a few subtypes with a clear promise towards improved performance once launched.

Accurate typing and mapping of these objects throughout the High Latitude Survey would allow for a detailed model of the structure of the Milky Way, the identification of lower contrast stellar streams and an accurate tally of the stellar halo. Taken over the Milky Way as a whole or per individual structural component, it would allow us to construct the low-mass end of the IMF, a critical component in understanding star and galaxy formation.